\documentclass[pdftex,twocolumn,epjc3]{svjour3}          

\RequirePackage[T1]{fontenc}

\smartqed 

\RequirePackage{graphicx}
\RequirePackage{mathptmx}      
\RequirePackage{flushend}
\RequirePackage[numbers,sort&compress]{natbib}
\RequirePackage[colorlinks,citecolor=blue,urlcolor=blue,linkcolor=blue]{hyperref}

\RequirePackage{amsmath}
\RequirePackage{amssymb}

\RequirePackage[usenames,dvipsnames]{xcolor}
\RequirePackage{slashed}

\RequirePackage{microtype}

\journalname{Eur. Phys. J. C}

\newcommand{\vphi}{\varphi}

\newcommand{\mbf}[1]{\mathbf{#1}}
\newcommand{\trm}[1]{\textrm{#1}}
\newcommand{\tsf}[1]{\textsf{#1}}

\newcommand{\be}{\begin{equation}}
\newcommand{\ee}{\end{equation}}
\newcommand{\bea}{\begin{eqnarray}}
\newcommand{\eea}{\end{eqnarray}}
\newcommand{\bi}{\begin{itemize}}
\newcommand{\ei}{\end{itemize}}

\newcommand{\nn}{\nonumber}
\newcommand{\Prob}{\tsf{P}}

\newcommand{\PBW}{\tsf{P}_{\scriptsize\tsf{BW}}}

\newcommand{\rmd}{\mathrm{d}}
\newcommand{\zmf}{\text{zmf}}
\newcommand{\arms}{a_\text{rms}}

\definecolor{aaaa}{rgb}{0.99, 0.4, 0.01}
\definecolor{bbbb}{rgb}{0.5, 0.3, 0.9}
\definecolor{grayEE}{rgb}{0.7, 0.7, 0.7}
\definecolor{bk1}{RGB}{230,40,40}

\newcommand{\figref}[1]{Fig. \ref{#1}}
\newcommand{\figrefa}[1]{Fig. \ref{#1}(a)}
\newcommand{\figrefb}[1]{Fig. \ref{#1}(b)}

\newcommand{\eqnref}[1]{Eq. (\ref{#1})}

\newcommand{\vkap}{\varkappa}
\newcommand{\cd}{.}

\begin{document}

\title{Higher fidelity simulations of nonlinear Breit-Wheeler pair creation in intense laser pulses}

\author{T. G. Blackburn\thanksref{e1,addr1}
\and 
B. King\thanksref{e2,addr2}}

\thankstext{e1}{e-mail: tom.blackburn@physics.gu.se}
\thankstext{e2}{e-mail: b.king@plymouth.ac.uk}

\institute{Department of Physics, University of Gothenburg, SE-41296 Gothenburg, Sweden\label{addr1}
          \and
          Centre for Mathematical Sciences, University of Plymouth, Plymouth, PL4 8AA, United Kingdom\label{addr2}
}

\date{Received: date / Accepted: date}

\maketitle

\begin{abstract}
When a photon collides with a laser pulse, an electron-positron pair can be produced via the nonlinear Breit-Wheeler process. A simulation framework has been developed to calculate this process, which is based on a ponderomotive approach that includes strong-field quantum electrodynamical effects via the locally monochromatic approximation (LMA).
Here we compare simulation predictions for a variety of observables, in different physical regimes, with numerical evaluation of exact analytical results from theory. For the case of a focussed laser background, we also compare simulation with a high-energy theory approximation.
These comparisons are used to quantify the accuracy of the simulation approach in calculating harmonic structure, which appears in the lightfront momentum and angular spectra of outgoing particles, and the transition from multi-photon to all-order pair creation.
Calculation of the total yield of pairs over a range of intensity parameters is also used to assess the accuracy of the locally constant field approximation (LCFA).
\end{abstract}

\section{Introduction}
A photon propagating through a laser pulse can decay to an electron-positron pair. The phenomenology of the interaction between the photon and the laser pulse is dependent on the centre of mass energy of the collision, the strength of the laser field, and the bandwidth of the pulse.
At weak field strengths, the leading-order contribution is determined by the number of photons that must be taken from the laser pulse to overcome the mass threshold.
If only one is necessary, the process is referred to as the linear Breit-Wheeler process~\cite{breit34}. At lower centre of mass energies, the leading-order kinematically-allowed process may involve several laser photons, which is the `multi-photon' nonlinear Breit-Wheeler process.
For higher field strengths, the `threshold' number of laser photons required for the photon to decay may not be the most probable channel, and all orders of interaction between the laser pulse and the photon must be taken into account.

To measure nonlinear Breit-Wheeler pair creation \cite{reiss62,nikishov64,narozhny69,heinzl10,PhysRevA.84.033416,PhysRevA.86.052104,PhysRevLett.108.240406,seipt12b,PhysRevA.88.062110,Titov:2013kya,Meuren:2014uia,Wu:2014zaa,Jansen:2015idl,Meuren:2015mra,Nousch:2015pja,PhysRevLett.117.213201,PhysRevD.94.013010,lobet.prab.2017,Hartin:2018sha,Seipt:2020diz,Ilderton:2019ceq,Ilderton:2019vot,Titov:2019kdk,Titov:2020taw,mercuri.njp.2021} in the `all-order' regime, a sufficiently powerful laser is required:
peak intensities $> 10^{22}~\text{W}\text{cm}^{-2}$ are now accessible with current laser technology~\cite{yanovsky.oe.2008,sung.ol.2017,kiriyama.ol.2018,yoon.optica.2021}.
Typical proposals involve providing a particle beam via laser-wakefield acceleration of electrons, which can produce energies of the order of several GeV \cite{kneip.prl.2009,wang.ncomm.2013,leemans.prl.2014,Gonsalves:2019wnc}.
This means that, in order for pair creation to be kinematically allowed, many photons are required. If the laser pulse is sufficiently intense, nonlinear Breit-Wheeler becomes sufficiently probable as to be measurable. In this regime of low energy and high field strength, the `locally constant field approximation' (LCFA) \cite{ritus85,harvey15,DiPiazza:2017raw,Ilderton:2018nws,DiPiazza:2018bfu,King:2019igt,Seipt:2020diz} of approximating the probability for the process as a sum of interactions with `instantaneously constant' phase slices of the laser pulse, is expected to be sufficiently accurate. The LCFA is the standard method for including strong-field QED effects in numerical simulation of laser-matter interactions \cite{Bell:2008zzb,Kirk:2009vk,nerush11,PhysRevSTAB.14.054401,ridgers14,PhysRevE.92.023305,Gelfer:2015ora,grismayer16}.

Recently, complementary `high-energy' experiments such as E320 at SLAC and LUXE \cite{Abramowicz:2021zja} at DESY have been suggested to measure all-order strong-field QED effects such as nonlinear Breit-Wheeler, Compton and the nonlinear trident process \cite{hu10,ilderton11,king13b,Dinu:2017uoj,King:2018ibi,Mackenroth:2018smh,Dinu:2019wdw,Torgrimsson:2020wlz}. These experiments  will collide particle beams accelerated using conventional radiofrequency cavities, with strong laser pulses. The particle beams can reach higher energies ($13$\,GeV for E320 and $11.5-16.5$\,GeV for LUXE), lower emittances, higher repetition rate and energy stability than current laser-wakefield accelerated beams have been achieved. These types of experiments therefore allow for: i) higher precision measurements of total and differential yields; ii) access to strong-field QED processes, such as nonlinear Breit-Wheeler, at lower intensities where the transition from perturbative to non-perturbative physics occurs. The field strengths where this transition takes place, are outside the region of applicability of the LCFA. Therefore, a new simulation framework must be developed to correctly model the physics at these experiments.

So far, the only measurement of the nonlinear Breit-Wheeler process was in the landmark E144 experiment, where, in the multiphoton regime \cite{hu10}, both nonlinear Compton scattering \cite{E144:1996enr} and Breit-Wheeler \cite{burke97} were measured. The experiment was modelled using an `instantaneously monochromatic' approximation \cite{Bamber:1999zt}, which has since been used in the simulation codes CAIN \cite{cain1} and IPStrong \cite{Hartin:2018egj}. This approximation has recently been formalised, by being derived directly from strong-field QED in a plane-wave background \cite{Heinzl:2020ynb}. The resulting form, the `locally monochromatic approximation' (LMA), assumes a pulse envelope that varies much slower than the carrier wavelength, and employs a local phase expansion, which includes interference effects between processes taking place within the same laser wavelength. Unlike the LCFA, the LMA is not restricted by its use in a particular intensity or energy regime, but it does assume that the interacting electromagnetic field is well-approximated by a plane electromagnetic wave.
Using a ponderomotive scattering approach, the LMA has been realised in the simulation code, Ptarmigan \cite{ptarmigan}, which is being used to model the interaction point physics of the LUXE experiment \cite{Abramowicz:2021zja}. A comprehensive benchmarking of the LMA with exact expressions from QED, was recently performed for the process of nonlinear Compton scattering \cite{Blackburn:2021rqm}, and an analysis of pulse shape effects beyond the LMA performed in \cite{King:2020hsk}.

Further motivation for improving the accuracy of modelling strong-field QED processes at the LUXE experiment, is provided by the search for new physics. The `LUXE New Physics search with Optical Dump' LUXE-NPOD \cite{Bai:2021dgm} is based on a secondary production mechanism that utilises the high-energy inverse Compton-scattered photons \cite{Seipt:2019yds} produced in the electron beam-laser collision, to produce axion-like particles (ALPs) in a beam dump further downstream of the experiment. Such a setup will be sensitive to ALPs with masses O(MeV)-O(GeV), which is a range that has attracted much attention in recent years \cite{Dolan:2014ska,Izaguirre:2016dfi,Marciano:2016yhf,Mariotti:2017vtv,Brivio:2017ije,Bauer:2017ris,Bauer:2018uxu,Hochberg:2018rjs,Csaki:2020zqz}.
It is important, therefore, that the modelling of the photon source is as accurate as possible.
The strong-field QED source of photons in LUXE-NPOD is modelled using Ptarmigan, which we further benchmark in the current paper. The approach we use to model strong-field QED in particle-laser collisions can be adapted to model the generation of ALPs at the interaction point itself \cite{Gies:2008wv,Villalba-Chavez:2014nya,Villalba-Chavez:2015xna,Dillon:2018ouq,Dillon:2018ypt,King:2018qbq,King:2019cpj}.

In the current paper, we calculate totally inclusive and differential probabilities for nonlinear Breit-Wheeler pair creation, using the LMA in a numerical simulation framework, and compare them to the prediction from strong-field QED employing plane-wave Volkov states \cite{volkov35}. Through a series of benchmarks for a typical pulse shape, for a range of parameters and observables, we acquire a measure of the accuracy of the LMA. Our comparison covers a regime soon to be explored in upcoming experiments, but also extends to much higher energies, where harmonic structure appears in particle spectra. We highlight pulse envelope effects that are beyond the LMA, and establish parameters where the LCFA can also be employed.
\newline

The paper is organised as follows. In Sec. 2, the phenomenology of nonlinear Breit-Wheeler is recapped, and the expressions used in the calculation directly from theory are stated; in Sec. 3 the implementation of the LMA in the simulation framework is explained; in Sec. 4 the results of benchmarking between the direct theory calculation and numerical simulation are presented. In Sec. 5 the results are discussed and the paper is concluded. Unless otherwise stated, $c$ and $\hslash$ have been set to unity.

\section{Theory background}
To aid the understanding of the rest of the paper, some of the main phenomenology associated with pair creation in a quasi-monochromatic field will be recapped. (Reviews of strong-field QED in a laser background can be found in \cite{ritus85,dipiazza12}.)
\newline

We define the rescaled vector potential $a=eA$, as:
\bea
a = m \xi \cos^{2}\left(\frac{\vphi}{2N}\right)\{0,\cos \vphi, \sin \vphi,0\}; \quad |\vphi| < N\pi \label{eqn:a1}
\eea
and $a=0$ otherwise, where $N$ is the number of cycles, \linebreak$\vphi=\kappa\cdot x$ is the phase, $\kappa$ is the laser wavevector (satisfying $\kappa\cdot a = 0$), $\xi$ is the classical nonlinearity parameter (also referred to as the `intensity parameter') and $m$ and $e>0$ are the positron mass and charge, respectively. We note that, using a vector potential that is non-zero only on a finite lightfront interval will mean having a background with a wide bandwidth. This will lead to an enhancement of pulse-envelope interference effects, which we will comment on when they arise.

The momentum contributed by the background field can be written in terms of harmonics $n$, of the central laser wavevector. In the LMA, conservation of momentum in nonlinear Breit-Wheeler can be written:
\be
    k + \bar{n} \kappa = q + q', \label{eqn:momcon}
\ee
where $q$ and $q'$ are the electron and positron quasimomenta, $q=p-(a^{2}/2\kappa\cdot p) \kappa $ and $p$ is the free electron momentum (and analogously $p'$ for the positron), $k$ is the momentum of the photon and we allow $\bar{n}$ to be a real number. Solving \eqnref{eqn:momcon} for $\bar{n}$, we find:
\bea
2\eta\bar{n} = \frac{1+\xi^{2}(\vphi)}{s(1-s)}+ \frac{s}{1-s}\left(\frac{\mbf{p}^{\perp}}{m}-\frac{1-s}{s}\frac{\mbf{p'}^{\perp}}{m}\right)^{2}, \nn \\
\label{eqn:pcons1}
\eea
where $s=\kappa\cdot p'/\kappa \cdot k$ is the lightfront momentum fraction, $\eta=\vkap\cdot k / m^{2}$ is the (photon) energy parameter and $\mbf{p}^{\perp}$ is the electron momentum transverse to the laser propagation direction (and analogously for the positron, $\mbf{p'}^{\perp}$). We have also defined $\xi^{2}(\vphi) = -a^{2}$, which, for the here-considered vector potential, \eqnref{eqn:a1}, becomes
    \begin{equation}
    \xi^{2}(\vphi) = \xi^{2} \cos^{4}\left(\frac{\vphi}{2N}\right); \quad |\vphi|<N\pi, \label{eq:LocalXi}
    \end{equation}
and $\xi(\vphi)=0$ otherwise. From \eqnref{eqn:momcon}, it can be seen $s\in [0,1]$ and since $\mbf{p}^{\perp}$, and $\mbf{p'}^{\perp}$ are integrated over, there is  no upper bound to $\bar{n}$. But there \emph{is} a lower bound, at $\mbf{p}^{\perp}=\mbf{p'}^{\perp}=0$ and $s=1/2$, giving:
\bea
\bar{n} \geq \frac{2[1+\xi^{2}(\vphi)]}{\eta}. \label{eqn:thresh1}
\eea
The `threshold' harmonic for pair creation to occur is then $n_{\star}=\lceil \bar{n} \rceil$. When $\xi \ll1$, the probability of this harmonic contributing to pair creation scales as $\sim\xi^{2n_{\star}}$, which is perturbative, because the next highest harmonic is a factor $\xi^{2} \ll 1$ compared to the leading harmonic, but $n_{\star}$ can be large and so the leading order generally depends nonlinearly on the background intensity, $\xi^{2}$. This is the `multi-photon regime'. As $\xi$ increases to a value where it no longer satisfies $\xi\ll1$, although one can still define a threshold harmonic, it is no longer the case that the standard harmonic hierarchy holds and that each successive higher harmonic is suppressed. Furthermore, the quasimomentum of the produced pair increases, and so it becomes more difficult to create a pair, and the curve of the probability arches down and away from the multiphoton scaling. If $\xi$ is increased sufficiently, the LMA tends to the LCFA (locally constant field approximation), which depends only on the combination of parameters, $\chi=\xi\eta$. At small $\chi$ (but large $\xi$) this allows for a tunneling dependency on $\chi$, whereas for large $\chi$, the probability for the tree-level process of nonlinear Breit-Wheeler scales as $\tsf{P}\sim \chi^{2/3}$. (This non-perturbative dependency on the field strength has recently been discussed in the context of the Ritus-Narozhny conjecture \cite{Fedotov:2016afw,Yakimenko:2018kih,Blackburn:2018tsn,Mironov:2020gbi,Ekman:2020fdp,Heinzl:2021mji,Torgrimsson:2021wcj,Berezin:2021sej}.) 

\begin{figure}
\centering
    \includegraphics[width=\linewidth]{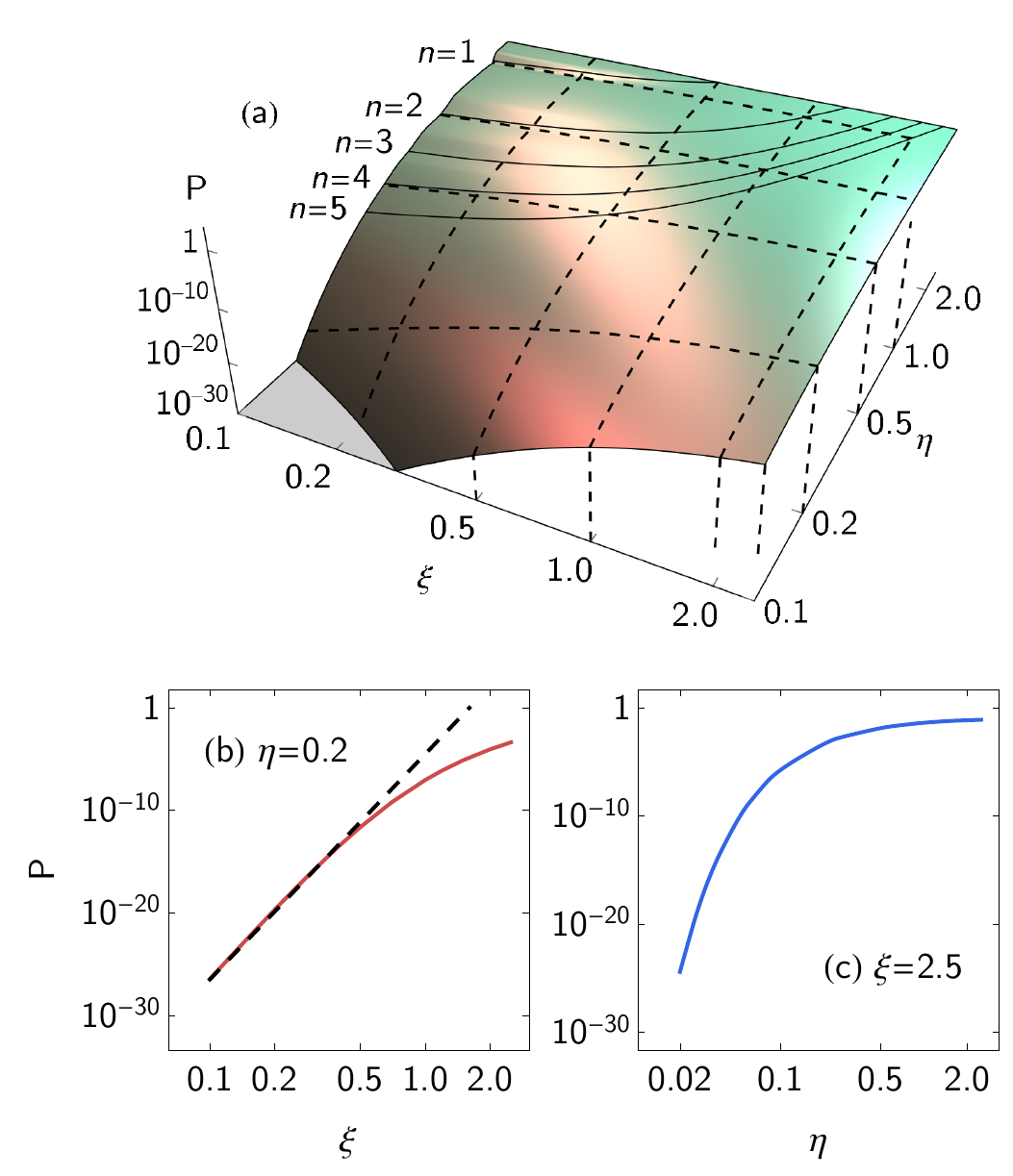}
\   \caption{%
    Demonstration of different parameter regimes in the probability of the nonlinear Breit Wheeler process, $\tsf{P}(\xi,\eta;N)$, for a $N=16$ cycle cos-squared pulse. The first five harmonic edges are drawn on the surface. Inset left: the multiphoton regime ($\xi \ll 1$), where at fixed $\eta$, $P\propto \xi^{2n_{\star}}$, where $n_{\star}$ is the threshold harmonic (for $\eta=0.2$, $n_{\star}=11$
    ). Inset right: the tunneling regime ($\xi \gg 1$, $\chi \ll 1$), in which the probability is exponentially suppressed as $\tsf{P}\sim \chi \exp(-8/3\chi)$.}
    \label{fig:surf1}
    \end{figure}

In a plane wave background, the total probability for pair creation, $\tsf{P}$, (or equivalently: yield of pairs per photon), will depend on the intensity parameter $\xi$, the probe particle energy parameter $\eta=\kappa\cdot k / m^{2}$ and the number of laser cycles $N$. We illustrate the various phenomenological regimes of $\tsf{P}(\xi,\eta,N)$ in \figref{fig:surf1}.

We calculate the probability for pair creation directly from QED in a plane-wave background. This involves using exact solutions to the Dirac equation in a plane wave potential (Volkov states \cite{volkov35}) to include the charge-laser coupling to all orders, and then perturbatively expanding in vertices of radiated particles. We evaluate the amplitude for each polarisation channel,  mod-square and then integrate over outgoing momenta. We outline the formulas used here for completeness and to introduce notation, but an analogous approach can be found in the literature. 

The total unpolarised probability $\Prob$ can be written as: 
    \begin{align}
    \Prob &= \frac{\alpha \mathcal{I}}{\eta},
    \label{eq:ProbabilityQED}
    \\
    \mathcal{I} &= \frac{1}{2}\sum_{\sigma_{k}=1}^{2}\sum_{\sigma_{p}=1}^{2}\sum_{\sigma_{p'}=1}^{2} \mathcal{I}_{\sigma_{k};\sigma_{p},\sigma_{p'}},
    \\
    \mathcal{I}_{\sigma_{k};\sigma_{p},\sigma_{p'}} &= \frac{1}{2^{4}\pi^{2}\eta}\int_{0}^{1}\frac{ds}{s(1-s)} \int \frac{d^{2}\mbf{r}^{\perp}}{m^{2}}|\tsf{S}_{\sigma_{k};\sigma_{p},\sigma_{p'}}|^{2},
    \end{align}
where $\mbf{r}^{\perp}=\mbf{p'}^{\perp}-s\mbf{k}^{\perp}$ and $\sigma_{k,p,p'}$
are the polarisation-state indices of the photon, electron and positron respectively. The (scaled) amplitude is given by:
\bea
\tsf{S}_{\sigma_{k};\sigma_{p},\sigma_{p'}}  &=& 
\int_{\vphi_{i}}^{\vphi_{f}}\!\!d\vphi ~\bar{u}_{\sigma_{p}}\tsf{S}_{\sigma_{k}}v_{\sigma_{p'}}\mbox{e}^{if} \nn \\
\tsf{S}_{\sigma_{k}}&=&\Delta \slashed{\epsilon}^{\ast}_{\sigma_{k}} + \frac{1}{2\,\kappa\cdot k}\left(\frac{\slashed{a}\slashed{\kappa}\slashed{\epsilon}^{\ast}_{\sigma_{k}}}{s}-\frac{\slashed{\epsilon}^{\ast}_{\sigma_{k}}\slashed{\kappa}\slashed{a}}{1-s}\right)
\eea
\bea
f &=&  \frac{1}{\eta(1-s)} \int^{\vphi}_{\vphi_i} \frac{k\cd \pi_{p'}(z)}{m^{2}}\,dz \label{eqn:Srprl1}
\eea
where  $\bar{u}_{\sigma_{p}}$ ($v_{\sigma_{p'}}$) are the outgoing electron (positron) bispinors,  $\Delta = 1-k\cd \pi_{p'}/k\cd p$ is the regularisation factor, $\pi_{p'}$ is the classical positron kinetic momentum:
\bea
 \pi_{p'} = p' + a - \kappa \left[\frac{p'\cd a}{\kappa\cd p'}+\frac{a\cd a}{2\kappa\cd p'}\right]. \label{eqn:pipp}
\eea
We will refer to the numerical evaluation of the above as the `QED' result. Although the formula is exact for a plane wave, the numerical evaluation will have a finite error.

It will be useful at times to compare our results with the probability of the linear Breit Wheeler process, $\PBW$, which is equal to the leading-order expansion in $\xi^{2}$ of \eqnref{eq:ProbabilityQED}:
\bea
\PBW = \frac{\alpha}{4\pi \eta^{2}}\int \frac{ds\,d(r^{2})}{s(1-s)} \left[h(s) + \frac{r^{2}}{(1+r^{2})^{2}} \right]|\mbf{\tilde{a}}(\psi)|^{2}, \label{eqn:pbw1}
\eea
where:
\[
\psi = \frac{1+r^{2}}{2\eta s(1-s)}; \qquad h(s)= \frac{1}{2s(1-s)}-1; \qquad r=\frac{|\mbf{r}^{\perp}|}{m}
\]
and $\mbf{\tilde{a}}(\psi)$ is the Fourier transform of the vector potential, $\mbf{a}(\psi) = \int d \vphi~\mbf{a}(\vphi) \exp(i\vphi \psi)$.

\section{Implementation in numerical simulations}

\subsection{Particle dynamics}

Numerical simulations of strong-field QED interactions are fundamentally semiclassical, in that probability rates for the processes under consideration are evaluated along the particles' classical trajectories~\cite{Blackburn:2019rfv,gonoskov21}.
The derivation of a `rate' from QED generally involves some level of approximation, as it must be locally well-defined.
Two such frameworks are available in our Monte Carlo simulation code, \textsc{ptarmigan}~\cite{ptarmigan}: the locally monochromatic~\cite{Heinzl:2020ynb} and locally constant field approximations~\cite{ritus85}.
In this work we focus on the LMA-based approach: for reviews of LCFA-based simulations, see \cite{ridgers14,gonoskov15}.
We have already described the simulation concept in \cite{Blackburn:2021rqm}, so we present only a summary here.

In the LMA approach, the classical trajectory is defined by the particle quasimomentum, which is the cycle-averaged value of the kinetic momentum.
For a photon, which propagates ballistically through the strong-field region, the quasi- and kinetic momenta coincide.
For an electron (or positron), the quasimomentum $q$ ($q'$), is the cycle average of the kinetic momentum (for the positron given in \eqnref{eqn:pipp}), and evolves according to the following equation of motion~\cite{quesnel.pre.1998}:
    \begin{align}
    \frac{\rmd}{\rmd \tau} q_\mu &= \frac{1}{2} m \partial_\mu \arms^2(X)
    \label{eq:PonderomotiveForce}
    \end{align}
where $\tau$ is the proper time and $\arms$ is the component of the potential that varies slowly with respect to the wavelength.
For a pulsed wave, we have $\arms(X) = \sqrt{\xi^2[\vphi(X)]}$, i.e. the square root of \eqnref{eq:LocalXi}.
The cycle-averaged position $X^\mu$, i.e. the component of the worldline that is slowly varying with respect to the laser wavelength, follows from $\rmd X^\mu/\rmd \tau = q^\mu / (m \sqrt{1 + \arms^2})$.
The local value of $q$ controls the probability rates of photon emission and electron-positron pair creation through the parameters $\arms$ and $\eta$: for an electron or positron, these are $\arms = \sqrt{q^2/m^2 - 1}$ and $\eta = \kappa . q /m^2$.

Using the LMA means that the fast oscillating component of the trajectory is included at the level of the probability rates, which incorporate the conservation of quasimomentum~\cite{Blackburn:2021rqm}.
In the LCFA framework, it is instead encoded in the trajectory itself, via the kinetic momentum, which evolves according to the Lorentz force equation:
    \begin{align}
    \frac{\rmd}{\rmd \tau} \pi_\mu &= \pm e F_{\mu \nu} \pi^\nu, & \pi^2 &= m^2
    \label{eq:LorentzForce}
    \end{align}
The rates of QED processes are then controlled by a single quantum parameter $\chi = e \sqrt{-(F_{\mu\nu} \pi^\nu)^2} / m^3$, which is defined instantaneously along the trajectory $x^\mu(\tau)$, and $\rmd x^\mu/\rmd \tau = \pi^\mu / m$.
As is done in the particle-in-cell approach to kinetic plasma simulations, the code uses ensembles of `macroparticles' to model real particle beams.
The number of real particles each macroparticle represents is called the `weight'.

\subsection{Event generation}

\paragraph{Under the LMA.}
The pair creation rate per unit lab time, $W$, of a photon with momentum $k$, embedded in a CP plane wave of normalized, root-mean-square amplitude $\arms$ and wavevector $\kappa$, is given by~\cite{Heinzl:2020ynb}
    \begin{multline}
    \frac{\rmd W_n}{\rmd s} =
        \frac{\alpha m^2}{k^0}
        \left\{
            J_n^2(z) -
            \frac{\arms^2}{2}
            \left[ \frac{1}{2 s (1 - s)} - 1 \right]
    \right. \\ \left. \vphantom{\frac{a^2}{2}}
            \times \left[ 2 J_n^2(z) - J_{n-1}^2(z) - J_{n+1}^2(z) \right]
        \right\}
    \label{eq:Rate}
    \end{multline}
where the argument of the Bessel functions, $z>0$ fulfils:
    \begin{equation}
    z^2 = \frac{4 n^2 \arms^2}{1 + \arms^2}
        \frac{1}{s_n s (1-s)}
        \left[ 1 - \frac{1}{s_n s (1-s)} \right],
    \end{equation}
and the auxiliary variables are
    \begin{align}
    s_n &= \frac{2 n \eta}{1 + \arms^2},
    &
    \eta &= \frac{\kappa.k}{m^2}.
    \end{align}
The lightfront momentum fraction $s = \kappa.q / \kappa.k$ is restricted to
    \begin{equation}
    \frac{1}{2} \left[1 - (1 - 4/s_n)^{1/2}\right] < s
    < \frac{1}{2} \left[1 + (1 - 4/s_n)^{1/2}\right],
    \label{eq:sLimits}
    \end{equation}
and therefore the rate is non-zero only for $s_n > 4$, or equivalently, harmonic orders $n > n_\star = \lceil 2(1+\arms^2)/\eta \rceil$.

A pair creation event occurs if, in a small interval of time $\Delta t$ along the trajectory (where $t$ is the lab time), the probability $W \Delta t$ satisfies $U < W \Delta t$, where $U$ is a pseudorandom number uniformly distributed in $(0,1)$.
In principle, the total rate $W$ can be obtained directly by integrating \eqnref{eq:Rate} over all $s$ and then summing all relevant $n$: however, it is much faster to precalculate $W$ as a function of $\arms$ and $\eta$ and implement the evaluation as a table lookup.
We restrict the sum to $n_\star \leq n \leq n_\text{max}$, where $n_\star$ is defined by \eqnref{eqn:thresh1}.
The cutoff $n_\text{max}$ is defined to be the lowest harmonic order that satisfies $W_n < 10^{-4} \sum_{i=n_\star}^{n} W_i$. In order to obtain a simple, analytical approximation for $n_\text{max}$ as a function of $\arms$ and $\eta$, we obtain it numerically for a range of $(\arms, \eta)$ points in the region $\arms < 10$ and $\eta < 1$ and then fit a trial function to the data obtained.
Our result is
    \begin{align}
    n_\text{max} &= n_\star + \lceil \Delta n \rceil,
    \\
    \Delta n &= \frac{0.25 (1 + 3.3 \sqrt{\arms} + 8.0\, \arms^2)(1 + 7.3 \eta)}{\eta}.
    \label{eq:SumLimits}
    \end{align}

We use the following procedure to obtain the quasimomentum of the created electron (equivalently, positron).
When pair creation occurs, the harmonic index $n$ and lightfront momentum transfer $s$ are pseudorandomly sampled from the emission rates, \eqnref{eq:Rate}.
The former is obtained by solving $U' = \sum_{i=n_\star}^n W_i / \sum_{i=n_\star}^{n_\text{max}} W_i$, where $U'$ is a pseudorandom number drawn on the unit interval and $W_n$ is the $n$th partial rate, i.e. \eqnref{eq:Rate} integrated over all $s$.
The cutoff, $n_\text{max}$, is the same that used when precalculating the total rate.
The lightfront momentum transfer $s$ is obtained by rejection sampling of \eqnref{eq:Rate}.

The momentum $q$ is fixed by $k$, $n$ and $s$.
In the zero momentum frame (ZMF), the electron (positron) is created with energy and momentum
    \begin{align}
    \epsilon_\zmf / m &= \sqrt{n \eta / 2},
    \\
    p_\zmf / m &= \left[ n \eta / 2 - (1 + \arms^2) \right]^{1/2}
    \end{align}
and scattering angle
	\begin{equation}
	\cos\theta_\zmf = (1 - 2 s) \epsilon_\zmf / p_\zmf
	\label{eq:CosThetaZMF}
	\end{equation}
The azimuthal angle $\varphi_\zmf$ is pseudorandomly determined in $0 \leq \varphi_\zmf < 2\pi$.
The momentum so obtained is then transformed back to the lab frame, using the fact that the four-velocity of the ZMF is $u_\zmf = (k + n \kappa) / \sqrt{(k + n \kappa)^2}$.
The quasimomentum of the other particle, $q'$, follows from quasimomentum conservation, $q' =  k + n \kappa - q$. 

\paragraph{Under the LCFA.}
We also compare the theoretical results to LCFA-based simulations.
The relevant rate, resolved in both positron energy $\varepsilon = f k^0$ and scattering angle $\vartheta$, is~\cite{baier94} (see also \cite{DiPiazza:2018bfu})
    \begin{equation}
    \frac{\rmd^2 W}{\rmd f \rmd z} =
        \frac{\alpha m^2 \zeta}{\sqrt{3} \pi k^0}
        \left[
            1 + \frac{f^2 + (1 - f)^2}{f (1 - f)}  z^{2/3}
        \right]
        K_{1/3}\!\left( \zeta z \right)
    \label{eq:LCFArate}
    \end{equation}
where the photon quantum parameter $\chi = e \sqrt{-(F_{\mu\nu} k^\nu)^2} / m^3$, $\zeta = 2/[3 \chi f(1-f)]$, $z = [2 (\varepsilon/m)^2 (1 - \beta \cos\vartheta)]^{3/2}$, velocity $\beta = \sqrt{1 - m^2/\varepsilon^2}$, and $K$ is a modified Bessel function of the second kind.
The domain of \eqnref{eq:LCFArate} is $0 < f < 1$ and $1 < z < \infty$.
If an electron-positron pair is created, the (kinetic) momenta are determined by conserving three-momentum $\vec{k} = \vec{\pi} + \vec{\pi}'$~\cite{duclous.ppcf.2011}, where the components of $\vec{\pi}'$ follow from the energy and angle pseudorandomly sampled from \eqnref{eq:LCFArate}.
Previous implementations have generally used the angularly integrated form of \eqnref{eq:LCFArate} and assumed that the electron, positron and photon momenta are all collinear, as $\vartheta$ is of order $m / k^0$.

\subsection{Biasing}

    \begin{figure}
    \centering
    \includegraphics[width=\linewidth]{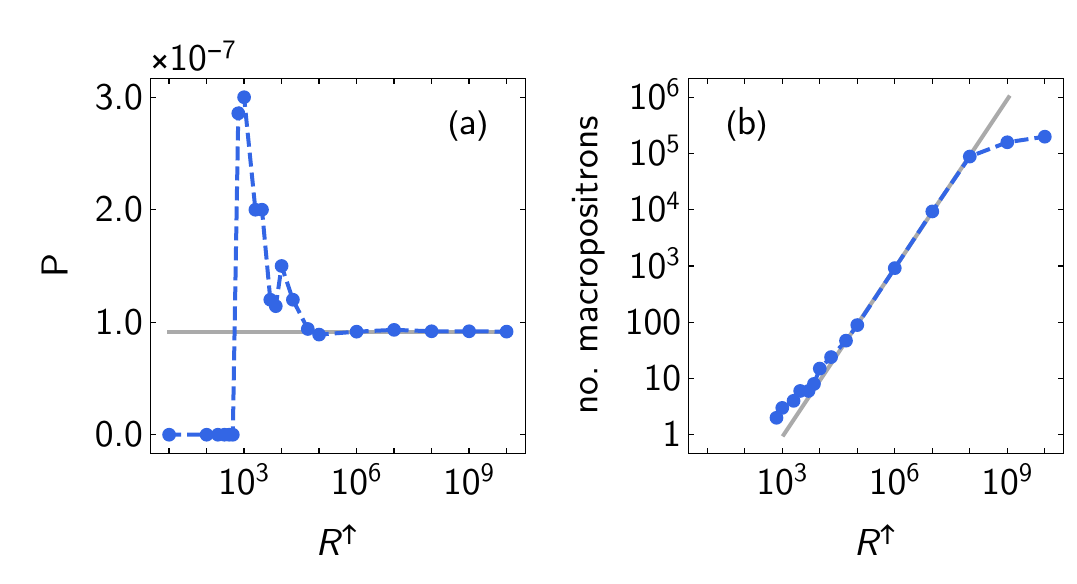}
    \caption{%
        (a) Pair creation probability and (b) number of macropositrons produced in Monte Carlo simulations of photons with energy parameter $\eta = 0.2$ colliding with a laser pulse of amplitude $\xi = 1.0$ and duration $N = 16$, as a function of the rate increase $R^\uparrow$.
        The number of macrophotons and their individual weights are fixed at $10^4$ and $10^{-4}$, respectively.
        Grey solid lines indicate the expected results, (a) $\Prob \simeq 10^{-7}$ and (b) $\text{no. macroparticles} = 10^4 \Prob R^\uparrow$.}
    \label{fig:RateIncrease}
    \end{figure}

The extreme rarity of pair creation events in certain regions of parameter space, which may be seen in \figref{fig:surf1}, is a challenge for Monte Carlo simulations.
As events are generated pseudorandomly, the number of macrophotons must be significantly larger than $1/\Prob \gg 1$ to resolve the positron yield (and larger still, to resolve differential quantities such as the spectrum).
In order to overcome this, our simulations implement a simple form of ``event biasing'', wherein the macrophoton partially decays.
The pair-creation rate, $W$, is artificially increased by a large factor $R^\uparrow \gg 1$, while the weights of any macroelectrons and macropositrons that are created are reduced by the same factor.
Thus, if the weight of the photon before pair creation is $w_\gamma$, the electron and positron are created with identical weights $w_\pm = w_\gamma / R^\uparrow$ and the photon weight is changed to $w'_\gamma = w_\gamma (1 - 1 / R^\uparrow)$.
As the code checks for pair creation in a single timestep $\Delta t$ by comparing the probability $R^\uparrow W \Delta t$ with a pseudorandom number drawn on the unit interval, it also sets an upper bound on $R^\uparrow$, such that the probability does not exceed 10\%.

We show an example of this procedure in action in \figref{fig:RateIncrease}.
The collision parameters ($\eta = 0.2$, $\xi = 1.0$ and $N = 16$) are such that the probability of pair creation $\Prob \simeq 10^{-7}$.
In the absence of event biasing, we would need $> 10^8$ macrophotons in order to resolve the yield with satisfactory accuracy.
Instead we use $10^4$ macrophotons, each with weight $w_\gamma = 10^{-4}$, and increase $R^\uparrow$ from 10 to $10^{10}$.
The probability, which is equivalent to the sum of the macropositron weights, is given in \figrefa{fig:RateIncrease} and the raw number of macropositrons in \figrefb{fig:RateIncrease}.
As may be expected, if $R^\uparrow \lesssim 10^4 \Prob \simeq 10^3$, no macropositrons are generated.
When $R^\uparrow$ increases above this value, the probability abruptly increases, fluctuates, and then converges to the expected value.
Convergence is driven by the increasing number of macropositrons, which grows linearly with $R^\uparrow$: the size of the fluctuations therefore scales as $1/\sqrt{R^\uparrow}$.
For the largest values of the rate increase, the number of macropositrons deviates from a linear scaling because of the upper bound that is automatically placed.

The simulation results we present in this manuscript are all generated using $R^\uparrow > 1$.
Choosing the value of the rate increase is a matter of balancing two competing concerns: on the one hand, if it is too small, the number of macropositrons is very small and the probability is poorly estimated; on the other hand, if it is too large, many low-weight particles must be advanced through the simulation domain and checked for, e.g., secondary photon emission.
Thus the code can spend an inordinate amount of time dealing with particles that have little overall importance to downstream uses such as detector simulation.
We have found that a reasonable balance is achieved by setting $R^\uparrow$ to the ratio between the photon-emission and pair-creation rates, evaluated at the same values of $\xi$ and $\eta$.

We conclude by noting that, in situations where the incident photon beam has a broad energy spectrum, increasing $R^\uparrow$ cannot compensate for poor statistics in the high-energy tail, which generally provides the dominant contribution to the positron yield.
In this case, the number of macrophotons must be increased as well.
In the present work, we deal exclusively with monoenergetic photon beams, so this is not a concern.

\section{Benchmarking}

In this section, we compare the predictions of numerical simulation, with direct evaluation of the QED expressions. Whilst the numerical simulations employ the LMA, the QED expressions do not. The aim is to understand how accurate a simulation employing the LMA can be, and in what parameter region it is accurate when calculating total yields and differential spectra. We expect agreement between the direct QED and LMA results everywhere that the LMA holds, with some disagreement for very short pulses (i.e. $N \not \gg 1$). In order to achieve this aim, we will first benchmark in a plane wave background,
where we know the direct QED expression to be accurate.

To benchmark in a plane wave background, we will compare the total yield, in each of the three variables: intensity, $\xi$; lightfront momentum $\eta$ and number of cycles, $N$. We will then benchmark some of the differential spectra: in positron lightfront momentum fraction, $s$ and in positron transverse momentum, $\mbf{r}^{\perp}$ (recalling that $\mbf{r}^{\perp}=\mbf{p'}^{\perp}-s\mbf{k}^{\perp}$) and $\mbf{k}^{\perp}$ is set to zero. As an example demonstrating a difference between QED and the LMA, we will consider a chirped laser pulse with more extreme parameters. Then to end this section, we will consider a focused background, which involves comparing two approximations with each other.

To quantify the accuracy of the simulation, we will define a measure of the error $\mathcal{E}$, given by:
\bea
\mathcal{E}_{\tsf{lma}}(\xi,\eta,N) = \frac{\tsf{P}(\xi,\eta,N)}{\tsf{P}_{\tsf{lma}}(\xi,\eta,N)}-1, \label{eqn:Edef}
\eea
where $\tsf{P}_{\tsf{lma}}$ is the probability as calculated from simulation using the LMA. (This form of the error was chosen so that it becomes large in the linear Breit-Wheeler region, where $P_{\trm{lma}}$ underestimates $P$.) We will also calculate $\mathcal{E}_{\tsf{lcfa}}(\xi,\eta,N)$, which is the equivalent accuracy measure, but where the locally constant field approximation is used in numerical simulation instead.

To make the comparison as faithful as possible, the simulation results have been generated with the recoil of electron/positron momenta due to photon emission, disabled.
Thus electrons and positrons that are created do emit secondary radiation, but their momenta are unchanged when they do so.
While recoil would not affect the yield, it would shift the momentum spectrum to smaller values of $s$.
Including these higher order effects, i.e. radiation reaction~\cite{DiPiazza:2010mv}, theoretically would involve a calculation of at least the `phototrident' process~\cite{morozov77,Torgrimsson:2020mto}, which is beyond the scope of the present work.

The intensity parameter range has been chosen to include where the LMA predicts something different to the standard approximation, the LCFA. Because it has been shown that in the region $\xi \gg 1$, $n_{\star} \gg 1$, the LMA tends to the LCFA in a plane wave background \cite{Heinzl:2020ynb}, and the LCFA has already been benchmarked in several works in the literature \cite{harvey15,DiPiazza:2017raw,Ilderton:2018nws,Blackburn:2018sfn,DiPiazza:2018bfu,King:2019igt}, we will concentrate on intensities that span the perturbative $\xi \ll 1$ to the `intermediate' intensity regime $\xi \sim O(1)$. 

The energy parameter range has been chosen to span the range from those that are just beyond experimentally accessible with state-of-the-art strong-field QED laser wakefield acceleration experiments \cite{cole18,poder18}, $\eta = 0.05$ (approximately $4\,\trm{GeV}$ for a head-on collision with optical laser photons), up to values where the harmonic structure becomes evident in spectra, $\eta \sim O(1)$. 

The pulse duration range has been chosen to reflect typically available laser pulse durations, with $N=16$ cycles (corresponding to a full-width-at-half-maximum duration of $20\,\trm{fs}$ for a wavelength of $1\,\mu\trm{m}$ [fs]) being a standard choice.

The pulse envelope shape is the one given in \eqnref{eqn:a1}: half a period of a squared cosine. Being a pulse with finite support has the advantage that it is more straightforward to benchmark with QED. However, such a pulse has a much wider Fourier spectrum than a pulse used in experiment (high and low energy tails of the spectrum would not be transmitted through all the optical elements). Therefore, we will see in what follows, that pulse envelope effects are significantly enhanced, but this allows us to better understand where the LMA may potentially fail.

\subsection{Pulsed plane waves}
\subsubsection{Yield}
\begin{figure}
    \centering
    \includegraphics[width=\linewidth]{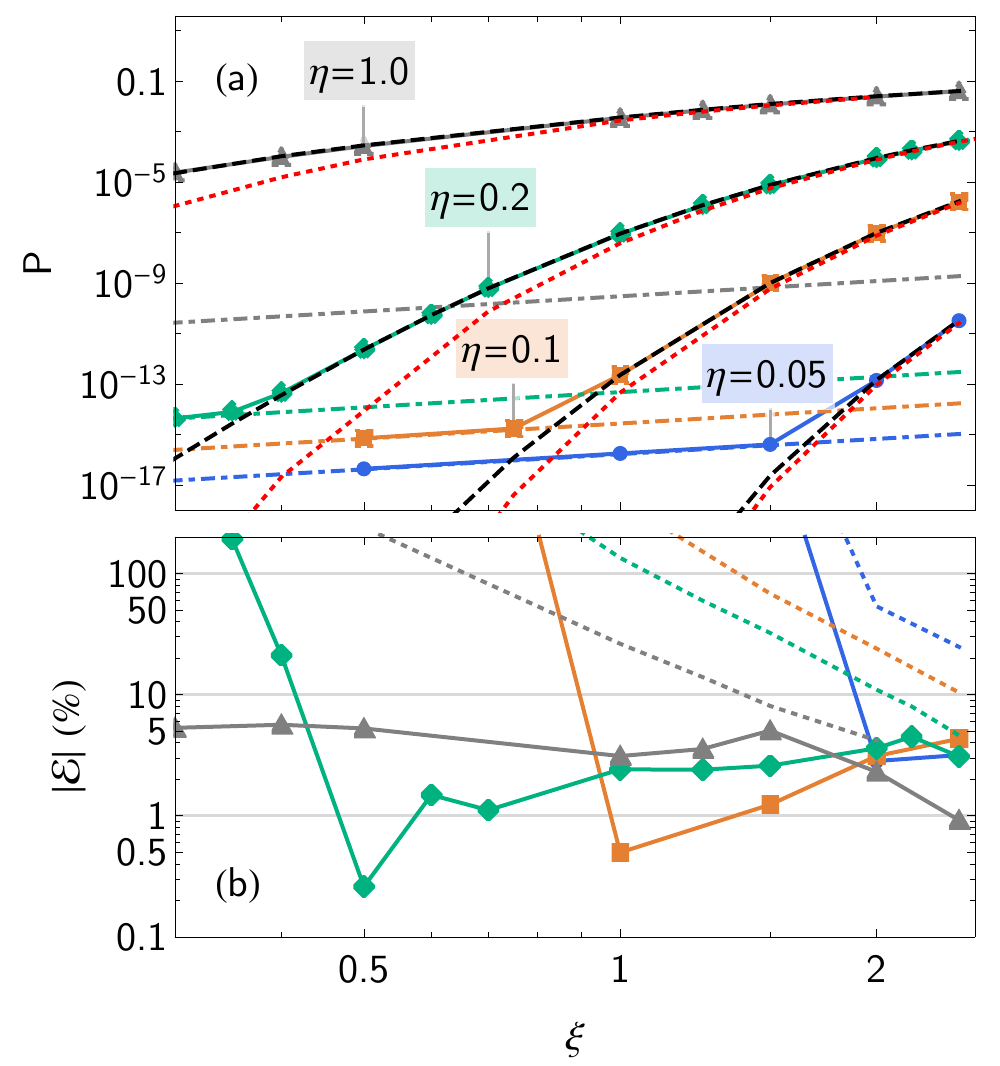}
    \caption{%
        (a) Pair creation probability $\Prob$ for a photon with energy parameter $\eta$ colliding with a circularly polarised plane-wave laser pulse of peak amplitude $\xi$ with $N = 16$ cycles. Results from theory are plotted for $\eta = 0.05$ (blue), $\eta = 0.1$ (orange), $\eta = 0.2$ (green) and $\eta = 1.0$ (grey), (all-orders: solid, perturbative: dot-dashed), as well as results from LMA-based simulations (black, dashed) and LCFA-based simulations (red, short-dashed).
        (b) Percentage error made by the simulations (LMA: solid, LCFA: dashed), calculated using \eqnref{eqn:Edef}.}
    \label{fig:yieldXi}
    \end{figure}

We refer to the totally inclusive probability as the \emph{yield}. First, we begin with benchmarking the yield as a function of the intensity parameter, $\tsf{P}=\tsf{P}(\xi;\eta,N)$ where $\eta$ and $N$ are constant values. In \figref{fig:yieldXi} we present results for $N=16$ which for a $1\,\mu\trm{m}$ wavelength (photon energy $1.24\,\trm{eV}$) laser background corresponds to a full-width-at-half-maximum duration of $20\,\trm{fs}$, and choose energy parameters $\eta\in\{0.05,0.1,0.2,1\}$, which correspond, for a head-on collision of a high-energy probe photon with the propagating wave, to photon energies of the order of $4, 8, 16, 80~\trm{GeV}$ respectively. We make the following observations.

For most of the intensity range, the LMA agrees very well with the direct QED result, with an approximate error of the order of $\mathcal{E} \approx 2-4\%$. We contrast this with the expansion parameter in the slowly-varying-envelope-approximation used in the LMA, which is of the order of $1/\Phi = 3.125\%$ for pulse duration $\Phi=2N$.

The LCFA is inaccurate for $\xi < 1$, which is expected, but we point out that due to the strong scaling with intensity in the multi-photon regime, the LCFA is incorrect by orders of magnitude. As is already known, the LCFA is more accurate for pair creation at higher photon energies, and this is reflected in the benchmarking. We also note the shallower gradient in how the accuracy of the LCFA increases as $\xi$ is increased: for example, an accuracy of $50\%$ is achieved at $\eta=1$ already when $\xi \approx 0.8$, but to achieve an accuracy of $5\%$ requires $\xi \approx 2.5$.

Below some value of the intensity parameter, $\xi$, the QED signal is dominated by the linear Breit-Wheeler process. A comparison is made in \figref{fig:yieldXi} between the QED result, and the linear Breit-Wheeler process from \eqnref{eqn:pbw1}. The enhancement of linear Breit-Wheeler is emphasised in the pulse form we have chosen. This is because the pulse is only non-zero for $|\vphi| < N\pi$, and hence is effectively multiplied by a flat-top envelope of width $2N\pi$. Since the linear Breit-Wheeler signal is proportional to the square of the Fourier transform of the potential, and since the flat-top envelope has a large bandwidth, this allows the linear Breit-Wheeler process to occur with a higher probability than e.g. in a Gaussian envelope of a comparable width.
proceed, compared to, e.g. a Gaussian envelope. Since this bandwidth effect, which is central to the linear Breit-Wheeler process, is due to pulse envelope interference effects, and since the LMA only includes carrier frequency interference, the LMA results deviate from the LMA at the linear Breit-Wheeler $\xi^{2}$ `floor' shown in \figref{fig:yieldXi}.

\begin{figure}
    \centering
    \includegraphics[width=\linewidth]{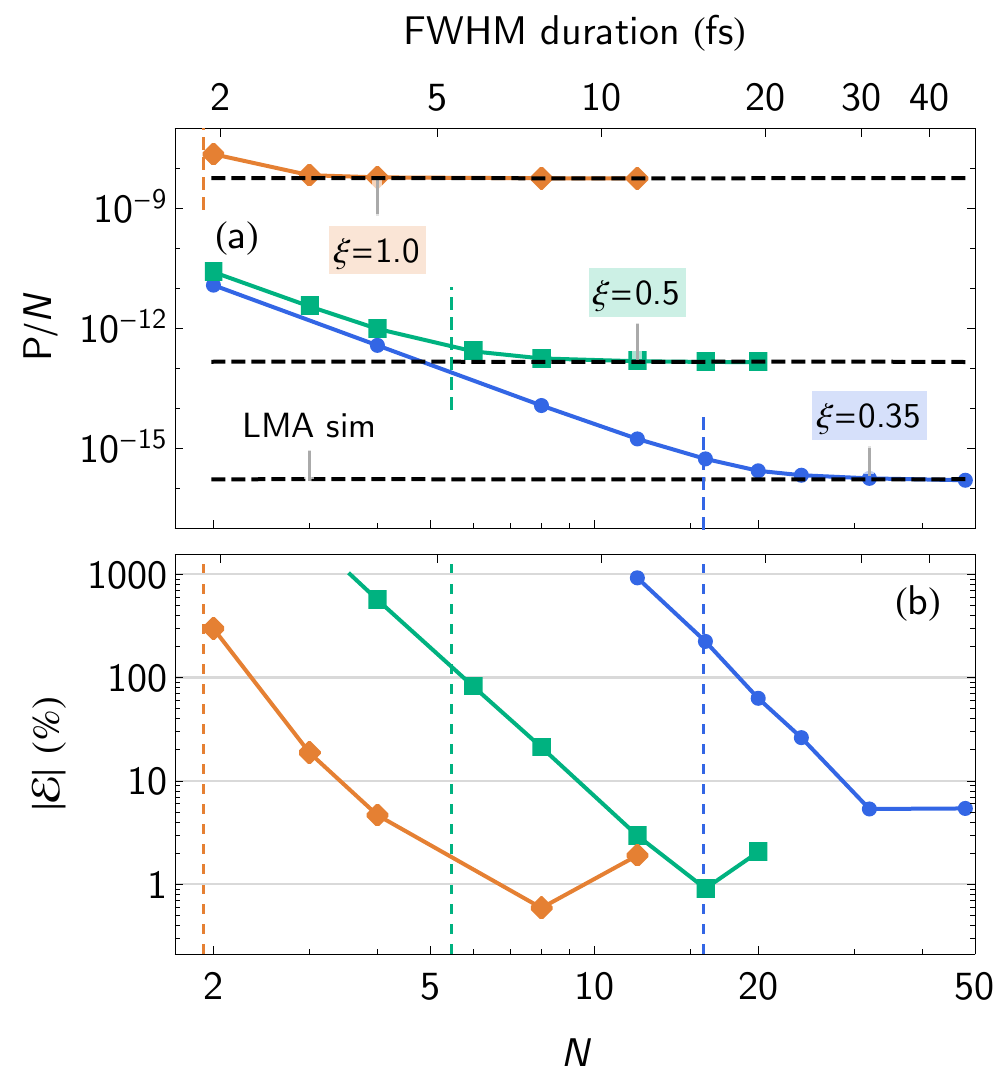}
    \caption{%
        (a) Pair creation probability $\Prob$ for a photon with energy parameter $\eta = 0.2$ colliding with a plane-wave, CP laser pulse of peak amplitude $\xi$ and duration equivalent to a number of wavelengths $N$: results from QED for $\xi = 0.35$ (blue), $\xi = 0.5$ (green) and $\xi = 1.0$ (orange) and LMA-based simulations (black, dashed).
        (b) Percentage error made by the simulations.
        Vertical, dashed lines give the $N$ at which the linear Breit-Wheeler contribution from the pulse envelope surpasses the multi-photon scaling from the LMA.}
    \label{fig:Duration}
    \end{figure}

To illustrate the dependency of the yield on the number of laser cycles, $N$, $\tsf{P}=\tsf{P}(N;\xi,\eta)$, we fix $\eta=0.2$ and choose $\xi$ from $\{0.35,0.5,1.0\}$. As we saw in \figref{fig:yieldXi}, for $N=16$, these $\xi$ values span a range including the linear Breit-Wheeler, multiphoton nonlinear Breit-Wheeler and nonperturbative nonlinear Breit-Wheeler physics. Since the bandwidth is proportional to $1/N$, we would expect that, as $N$ is reduced, the contribution from linear Breit Wheeler should increase. In \figref{fig:Duration}, this is indeed what we find. This is contrasted with the prediction from the LMA. The LMA probability depends on the pulse envelope, which is a function of $\vphi/2N$ and so a simple change of integration variables shows that the LMA simply scales linearly with $N$, which is also demonstrated in \figref{fig:Duration}.

One can estimate the pulse parameters where the linear contribution dominates by equating the perturbative result \eqnref{eqn:pbw1}, with the prediction from the LMA. The predicted value of $N$ where linear Breit Wheeler is of equal probability to the LMA prediction, is indicated by the vertical, dashed lines in \figref{fig:Duration}, which shows good agreement with the QED results.
\newline

\begin{figure}
    \centering
    \includegraphics[width=\linewidth]{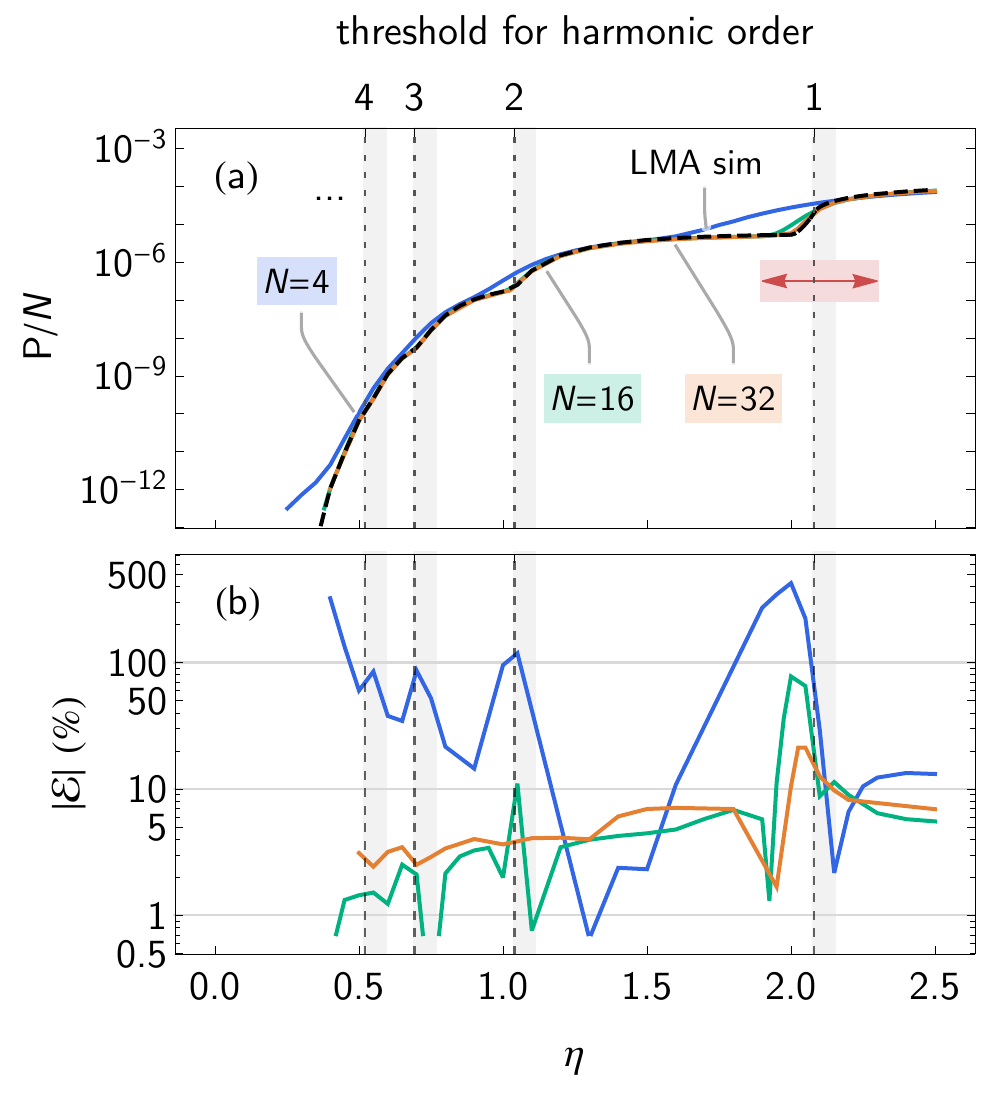}
    \caption{%
        (a) Pair creation probability $\Prob$ for a photon with energy parameter $\eta$ colliding with a circularly-polarised plane-wave of peak amplitude $\xi = 0.2$ and duration equivalent to a number of wavelengths $N$. Results are plotted from QED for $N = 4$ (blue), $N = 16$ (green) and $N = 32$ (orange) and LMA-based simulations (black, dashed).
        (b) Percentage difference between simulation and direct QED calculations.
        Vertical, grey dashed lines give the threshold in $\eta$ to access an additional harmonic channel.
        The change in the spectrum in the region around the $n = 1$ channel crossing, highlighted in red, is shown in detail in \figref{fig:StaircaseSpectra}.}
    \label{fig:Staircase}
    \end{figure}

To complete this yield section, we analyse the dependency on lightfront momentum, $\eta$. In \figref{fig:Staircase}, we plot the function $\tsf{P}=\tsf{P}(\eta;\xi,N)$, fixing the field parameter at $\xi_{0}=0.2$, and comparing three different fixed number of laser cycles, $N_{0}\in\{4,16,32\}$. We recall that the `threshold harmonic', $n_{\star}=\lceil \bar{n} \rceil$ (with $\bar{n}$ given by \eqnref{eqn:thresh1}), is a function of phase, and therefore in a given laser pulse, different parts of the pulse can access different threshold harmonics. However, we also recall that the higher the field parameter, $\xi$, the more probable the Breit-Wheeler process is. When considering the yield of pairs the relevant threshold harmonic, is the one given by the maximum value of $\xi(\vphi)$ in the pulse, namely $\xi$. In \figref{fig:Staircase}, we note that the $\eta$ values of the first four harmonics, correspond to `steps' in the  `staircase'-like dependency of the yield on $\eta$. For the short pulse ($N=4$), the full QED result gives a smooth increase at each harmonic, and as the pulse increases in duration, the QED results tends towards the LMA prediction from simulation. Since the LMA involves application of the `slowly-varying-envelope-approximation' and neglects terms in the Kibble mass of order $\sim 1/N$, we expect better agreement with QED as $N$ is increased, which is indeed what we find. 

In \figref{fig:Staircase}(b), we plot the error function, \linebreak $\mathcal{E}(\eta;\xi,N)$. Whilst the general trend is that a longer pulse (larger $N$), leads to a smaller value of $\mathcal{E}$, if the parameters are such that $\tsf{P}$ is close to a channel opening, the error can increase. This is particularly the case for shorter pulses, where the slowly-varying-envelope-approximation is already predicted to lead to a lower accuracy. However, we see that if the pulse is sufficiently long, $\mathcal{E}$ becomes relatively insensitive to the channel-opening effect and the error saturates. 

Finally, we note how \figref{fig:Staircase} shows that if $\eta$ is reduced enough, the probability due to the carrier frequency drops sufficiently, that the finite bandwidth linear Breit-Wheeler probability, contained in the QED but not the LMA, dominates again.

\subsubsection{Lightfront momentum spectra}

    \begin{figure*}
    \centering
    \includegraphics[width=0.8\linewidth]{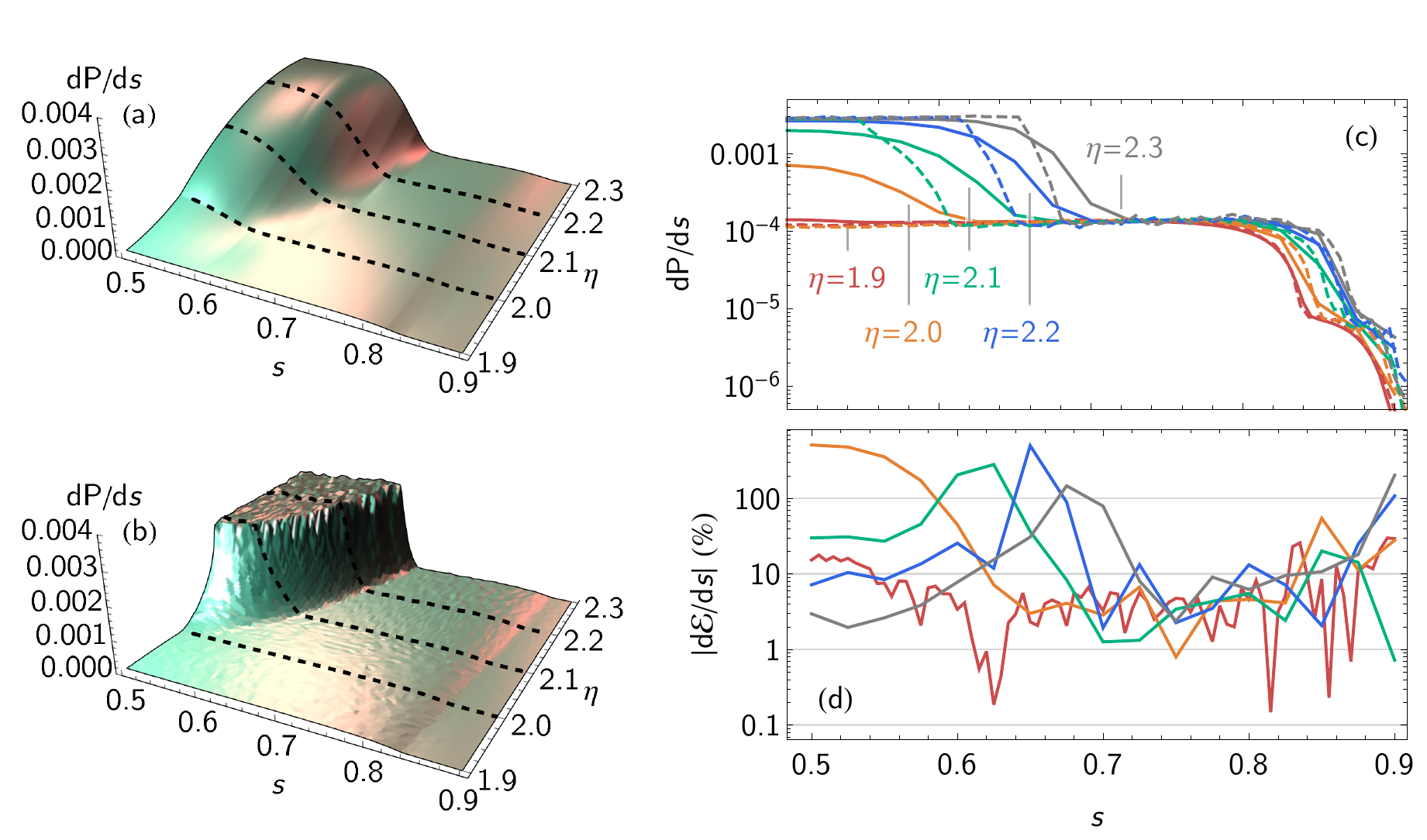}
    \caption{%
        Positron lightfront-momentum spectra $\rmd\Prob/\rmd s$ around the $n = 1$ channel crossing,
        for a laser pulse with peak amplitude $\xi = 0.2$ and $N = 16$:
        (a) results from QED and (b) LMA-based simulations.
        (c) Spectra at specific values of $\eta$, from QED (solid) and simulations (dashed) from the energy parameter region indicated in \figrefa{fig:Staircase}.
        (d) The percentage error.}
    \label{fig:StaircaseSpectra}
    \end{figure*}

Here we plot the single differential probability, \linebreak $d\Prob(s;\xi,\eta,N)/ds$, which, when integrated over $s\in[0,1]$, gives the total yield, $\Prob$, as in the previous section. We pick two cases to investigate:
i) $\xi=0.2$, $N=16$, and various energy parameters $\eta \in [1.9,2.4]$, which correspond to the opening of the $n=1$ harmonic channel; and
ii) $\eta=0.2$, $N=16$, and various intensity parameters around the intermediate range, $\xi \in [0.5,2.5]$.
The parameters in i) demonstrate how the harmonic channel-opening phenomenon is approximated by the LMA; those in ii) correspond to those that will be used in upcoming experiments \cite{Abramowicz:2021zja}, but for which there is less structure in the spectra as the threshold harmonic is much larger than one. Due to the symmetry of the spectrum around $s=1/2$, we will only plot the $1/2 \leq s < 1$ part of the spectra.

The plot of the yield in \figref{fig:Staircase} revealed a staircase-like structure due to the opening of harmonic channels as $\eta$ is varied. This phenomenon can also be revealed in the lightfront momentum spectrum. In \figref{fig:StaircaseSpectra}, we demonstrate this, by picking parameters that best display the effect, but are beyond what has been considered for near-future experiments. The parameters $\xi=0.2$ and $N=16$ are set as constant, and various constant values of $\eta$ are plotted. For these parameters, the first harmonic channel `opens' when $\eta=2.08$. Therefore in \figref{fig:StaircaseSpectra}, we vary $\eta$ parameters across this value, to show how channel-opening manifests in the lightfront momentum spectrum. We find that, as $\eta$ is increased, the centre of the distribution rises as a peak, demonstrating channel opening first, and as $\eta$ is further increased, the width of the peak broadens. Recalling the introductory discussion about momentum conservation and the results  \eqnref{eqn:pcons1} and \eqnref{eqn:thresh1}, we note that the `threshold' harmonic corresponds to a condition fulfilled by the most probable lightfront momentum, which is at $s=1/2$. The `threshold' then increases as $s$ is increased/decreased from the central value. The result in \figref{fig:StaircaseSpectra} reflects this behaviour.

Comparing the QED with the LMA result, we see that whereas QED describes a smooth transition as the harmonic range is crossed over, the LMA result displays a much clearer `jump'. For $\eta<2.08$, there is no discernible change in the LMA spectrum, which reflects the fact that the LMA is only including interference effects over a wavelength of the pulse, and therefore the transition is not `softened' by the finite spectrum of the pulse envelope. This `jump' of the LMA, is demonstrated more clearly in \figref{fig:StaircaseSpectra}, where the simulation and QED results are compared with $\eta$ as a parameter as well as $s$. The sudden jump of the simulation spectrum can clearly be seen, although at $s=1/2$, some smoothness can be discerned. Also visible in \figref{fig:StaircaseSpectra} is the second harmonic, which is sharper in the simulation results than in QED. Also evident, is the Monte-Carlo noise on the simulation results.

    \begin{figure}
    \centering
    \includegraphics[width=\linewidth]{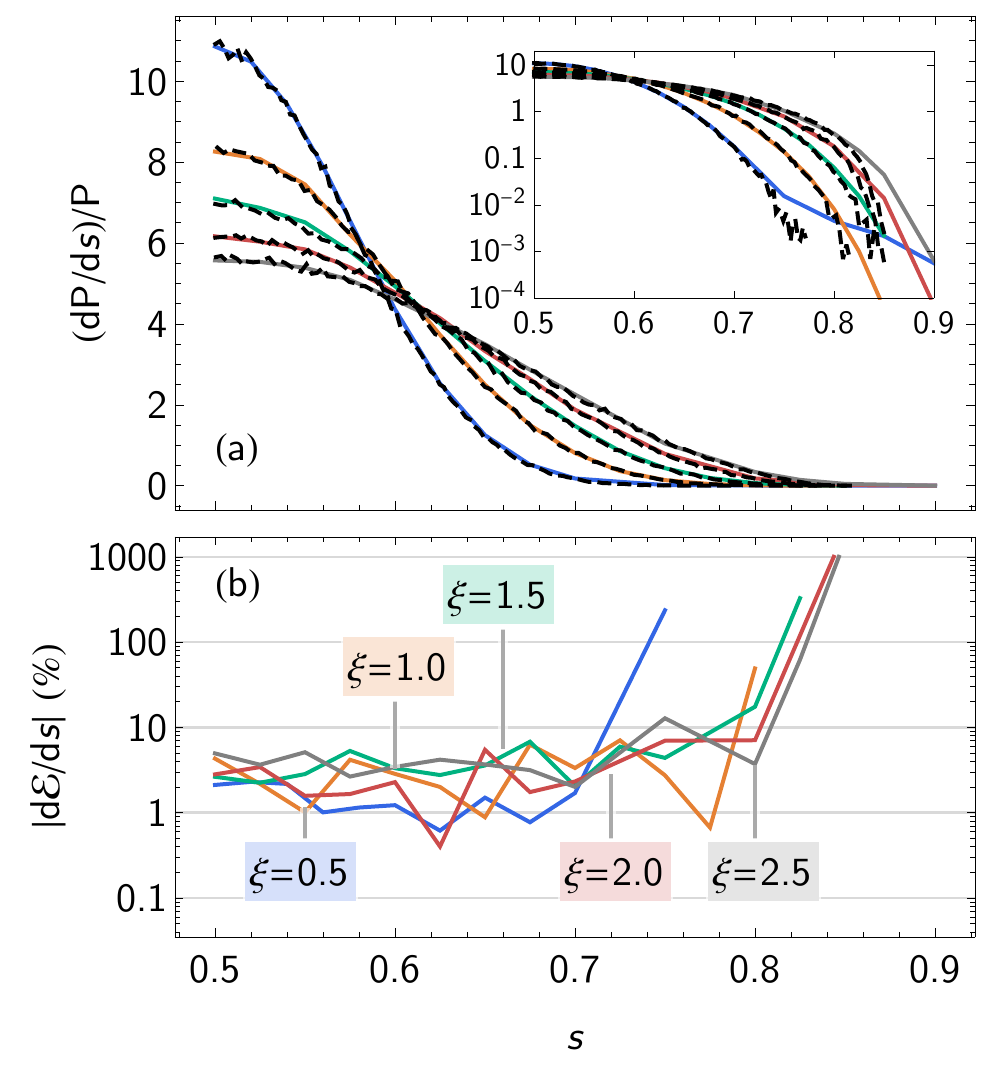}
    \caption{%
        (a) Lightfront momentum spectra $\rmd\Prob/\rmd s$ for a photon with energy parameter $\eta$ colliding with a plane-wave, CP laser pulse of peak amplitude $\xi$ and duration equivalent to a number of wavelengths $N = 16$: results from QED for $\xi = 0.5$ (blue), $\xi = 1.0$ (orange), $\xi = 1.5$ (green), $\xi = 2.0$ (red) and $\xi = 2.5$ (grey); and LMA-based simulations (black, dashed).
        (b) Percentage error made by the simulations.}
    \label{fig:tspecXi}
    \end{figure}

We illustrate the spectra for the soon experimentally-accessible case, in \figref{fig:tspecXi}. To aid comparison, the spectra have been normalised by their maximum values. We note the behaviour, that as the intensity increases, the lightfront-momentum spectrum becomes wider. Calculation of the first differential of the error function, $d\mathcal{E}/ds$ reveals an accuracy of around the $2-5\%$ level for all the intensity values, with the exception of $\xi=0.5$. We recall from the study of the yield \figref{fig:yieldXi}, that $(\xi,\eta,N)\approx(0.5,0.2,16)$ is the point at which the LMA started to diverge from QED due to the linear Breit-Wheeler effect for the parameters we have chosen. Since linear Breit-Wheeler has a wider momentum spectrum \cite{king21dd}, and since at $\xi=0.5$, the spectrum is heavily suppressed at values of momentum in the `tails', if $s$ is far enough from the centre at $s=1/2$, the linear Breit-Wheeler contribution will dominate. This explains the large increase in the error in \figref{fig:StaircaseSpectra} for $s>0.7$ and $\xi=0.5$.
    
    \begin{figure*}
    \centering
    \includegraphics[width=0.8\linewidth]{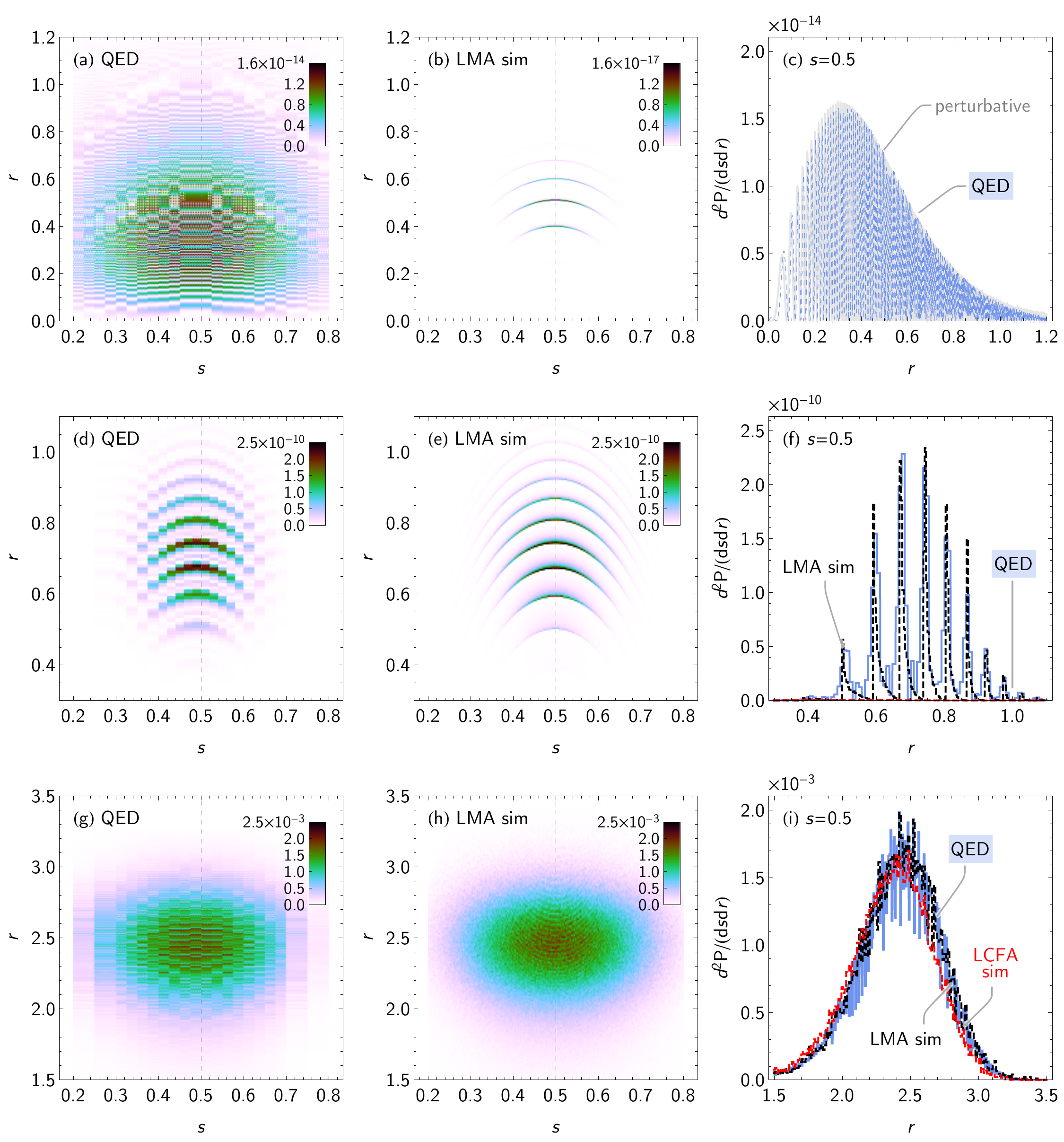}
    \caption{%
        Double-differential spectra (lightfront momentum $s$ and scaled perpendicular momentum $q_\perp$) for positrons produced by a photon with energy parameter $\eta$ in a laser pulse with peak amplitude $\xi = 0.2$ (top row), $\xi = 0.5$ (middle row), $\xi = 2.5$ (bottom row).
        Left column: QED results.
        Centre column: simulation results.
        Right column: Lineouts at $s = 0.5$.}
    \label{fig:angPlot}
    \end{figure*}
    
\subsubsection{Angular spectra}
In this section, the double differential spectrum in the \linebreak positron's perpendicular momentum $r$ (recalling that $r=|\mbf{r}^{\perp}|/m$) and the the lightfront momentum, $s$, is presented. In calculating $d^{2}\tsf{P}(r,s;\xi,\eta,N)/dr ds$, $\eta=0.2$ and $N=16$ are chosen constant, and three cases of constant $\xi$ are plotted. $\xi=0.2$ has been chosen as an example of the linear Breit-Wheeler signal, $\xi=0.5$ is in the multiphoton regime, and $\xi=2.5$ is in the all-order regime.

In the $\xi=0.2$ plot in \figref{fig:angPlot} (top row), the \linebreak  double-differential spectrum has a fine, highly-oscillating structure. The period of oscillation becomes shorter as $r$ is increased from $r=0$. In the figure, above $r\approx 0.3$, the oscillation is no longer resolved by the density of data points that was calculated, and any structure along the $r$-axis that can be discerned above this value, is just an aliasing effect. A lineout at $s=0.5$ is presented for comparison. When the corresponding spectrum is calculated using the linear Breit-Wheeler formula \eqnref{eqn:pbw1}, we see a good agreement when $r<0.3$, but where the full QED spectrum is slightly shifted with respect to the perturbative result. Although $r\gtrsim 0.3$ is difficult to resolve, the overall shape and magnitude of the rest of the spectra show excellent agreement. For comparison in the middle plot, the prediction of the LMA is presented, but as this only includes carrier-frequency and not pulse-envelope interference, misses the linear Breit-Wheeler contribution entirely, as expected.

The $\xi=0.5$ plots in \figref{fig:angPlot} (middle row), illustrate the double-differential spectrum in the multi-photon case. The threshold harmonic is $n_{\star}=13$, but already at $\xi=0.5$, the threshold harmonic does not give the leading contribution, rather this occurs at some orders above the threshold. In the figure it is the harmonics from $n=15$ to $n=23$ harmonics that are visible in the spectrum. In the lineout at $s=0.5$ (right-hand plot), the main contribution from the QED result agrees very well with the LMA from numerical simulation. Sub-harmonics between the main peaks, which are due to interference on longer length scales than a wavelength (and hence are beyond the LMA), can also be observed in the QED result.

The $\xi=2.5$ plots in \figref{fig:angPlot} (final row) demonstrate the angular spectra in the 'all-order' regime, where many harmonic orders contribute to pair creation. Although the threshold harmonic here is $n_{\star}=73$, by the number of peaks in the plots, it can be seen that the harmonic order that contributes most, is much larger than $n=73$. In the parameter region $n\gg 1$, $\xi \gg 1$, it is known that predictions using the LMA tend to those using the LCFA \cite{Heinzl:2020ynb}. Confirmation of this is seen in the right-hand panel lineout at $s=0.5$, there the LCFA shows good agreement with both the QED and the LMA curves. (This agrees with what we expect from the yield plot in \figref{fig:yieldXi}.)

\subsection{Chirped pulses}
\begin{figure}[h!!]
    \centering
    \includegraphics[width=\linewidth]{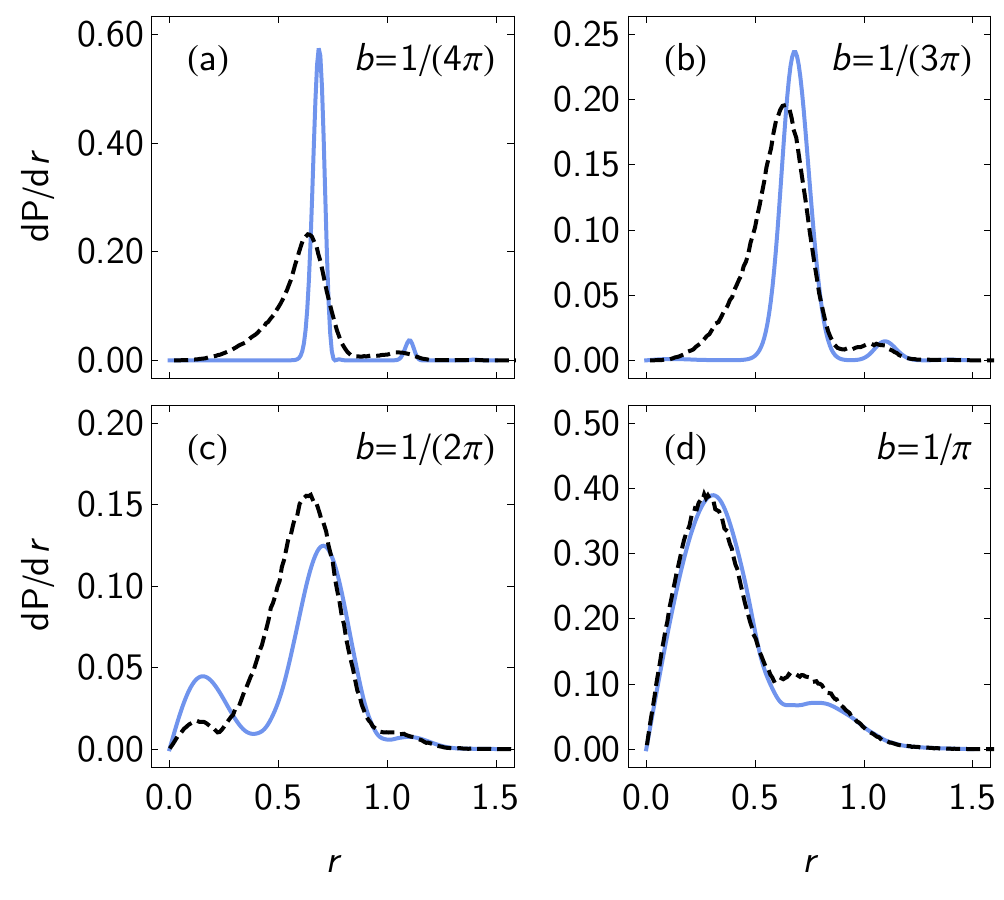}
    \caption{%
        Perpendicular momentum spectra $\rmd\Prob/\rmd r$ for a photon with energy parameter $\eta = 1.5$ colliding with a plane-wave, CP laser pulse of peak amplitude $\xi = 0.2$, $N = 16$ and chirp parameter $b$:
        results from QED (blue, solid) and from LMA-based simulations (black, dashed).}
    \label{fig:chirp}
    \end{figure}

The LMA can be used with a spacetime-dependent wavevector. An example of this is a chirped plane wave. In this section, we consider pair creation in a symmetrically chirped pulse with scaled vector potential defined over the domain $|\vphi| < N\pi$ as:
\[
a = m \xi \cos^{2}\left(\frac{\vphi}{2N}\right)\{0,\cos [\psi(\vphi;b,N)], \sin [\psi(\vphi;b,N)],0\},
\]
with the chirp function:
\bea
\psi(\vphi;b,N) = \vphi-\frac{b\,\vphi^{2}}{2N} \label{eqn:achirp}
\eea
and $a=0$ otherwise. This would give a `local' energy parameter of the form $\bar{\eta}(\vphi) = \eta(1-b\vphi/N)$, where $\eta$ is the `unchirped' value. At the centre of the pulse, $\bar{\eta}=\eta$ and the chirp is effectively zero, but towards the leading edge of the pulse $\bar{\eta}$ takes its maximum value. In \figref{fig:chirp}, a comparison is made between QED and simulation, for the calculation of the transverse momentum spectrum, $d\tsf{P}(r;\xi,\eta,N,b)/dr$, with $\xi=0.2$, $\eta=1.5$, $N=16$ and the chirp parameter $b$ varied up to its highest, physical value $b=1/\pi$ (at which point, at the trailing of the pulse, $\bar{\eta}(\vphi)= 0$). This set of parameters is a rather extreme case: in the middle of the pulse, $\bar{\eta}(0)=1.5$, which is almost midway between the energy values corresponding to the $n=2$ and $n=1$ harmonics (as can be seen from \figref{fig:Staircase}). The leading edge of the pulse is at a frequency which is below the $n=1$ harmonic threshold  \figref{fig:chirp} for $b=1/4\pi$ and $b=1/3\pi$ (first row), and above the harmonic threshold for $b=1/2\pi$ and $b=1/\pi$ (second row). Therefore the parameters describe a channel opening due to \emph{chirp} effects, where, in different parts of the pulse, the probe photon can access different threshold harmonics.  
From studying channel opening by varying the energy of the photon in \figref{fig:StaircaseSpectra}, we expect the errors of the LMA to be larger in this case because it predicts a more sudden channel opening than in QED, as $\eta$ is varied. In \figref{fig:chirp} we indeed see that, at $b=1/4\pi$, the QED result, which demonstrates sub-threshold pair creation, is noticeably different to the LMA prediction. However, as the chirp is increased, and the LMA describes pair creation above the $n_{\ast}=1$ threshold, the agreement between QED and simulation improves.

\subsection{Focused lasers}
In any real experiment, finite-size effects such as the laser pulse waist and focusing will be important in determining the yield of pair-creation events and in influencing particle spectra. Although no analytical form of an electron wavefunction in a typical focused laser background is yet known (but see e.g. \cite{Heinzl:2016kzb,Heinzl:2017zsr,Heinzl:2017blq}), we can employ a high-energy approximation \cite{DiPiazza:2013vra,DiPiazza:2015xva,DiPiazza:2016maj} that averages over the probability for the process in a plane-wave background:
\bea
\tsf{P}_{\tsf{3D}} = \int d^{2}\mbf{x}^{\perp}\,\rho(\mbf{x}^{\perp})\Prob[a(\mbf{x}^{\perp})], \label{eq:ProbQED2D}
\eea
where $\mbf{x}^{\perp}$ is the spatial co-ordinate transverse to the laser propagation direction, $\rho(\mbf{x}^{\perp})$ is the photon probe areal density and the plane-wave probability $\Prob$ is calculated from \eqnref{eq:ProbabilityQED}. The expression in \eqnref{eq:ProbQED2D} is then taken as the `QED result' (although it also involves an approximation) and compared with simulation employing the LMA.

The paraxial Gaussian beam \cite{salamin_review06} is taken to model the focused laser pulse, which has a potential of the form:
\bea
\mbf{a}^{\perp} = \frac{m\xi g(\vphi)}{\sqrt{1+\varsigma^{2}}} \exp\left[-\frac{(\mbf{x}^{\perp})^{2}}{w_{0}^{2}(1+\varsigma^{2})}\right][\cos \psi(\vphi),\sin \psi(\vphi)]\nn \\
\eea
where $g(\vphi)$ is a pulse envelope, $\varsigma=z/z_{r}$, the Rayleigh range $z_{r}=\omega w_{0}^{2}/2$, the carrier frequency of the laser pulse is $\omega$, the beam waist is $w_{0}$ and the phase dependency is:
\[
\psi(\vphi) =\vphi - \frac{(\mbf{x}^{\perp})^{2}\varsigma}{1+\varsigma^{2}}+\tan^{-1}\varsigma,
\]
with $\vphi=\omega(t-z)$. In order to use the high-energy approximation \eqnref{eq:ProbQED2D} for the QED calculation with the paraxial Gaussian beam, we employ the infinite Rayleigh-length approximation \cite{Gies:2017ygp,King:2018wtn} (no such approximation is required in the simulation), which sends $\varsigma\to0$, meaning that $\mbf{a}^{\perp}$ can be written as $\mbf{a}^{\perp} = \exp[-(\mbf{x}^{\perp})^{2}/w_{0}^{2}]\mbf{a}^{\perp}_{\tsf{pw}}$ where $\mbf{a}^{\perp}_{\tsf{pw}}$ is the plane-wave potential. We pick the same pulse envelope as in previous sections, i.e. $g(\vphi) =\cos^{2}\left(\frac{\vphi}{2N}\right)$ for $|\vphi|<N\pi$ and $g(\vphi)=0$ otherwise.

\begin{figure}[h!!]
\centering
    \includegraphics[width=\linewidth]{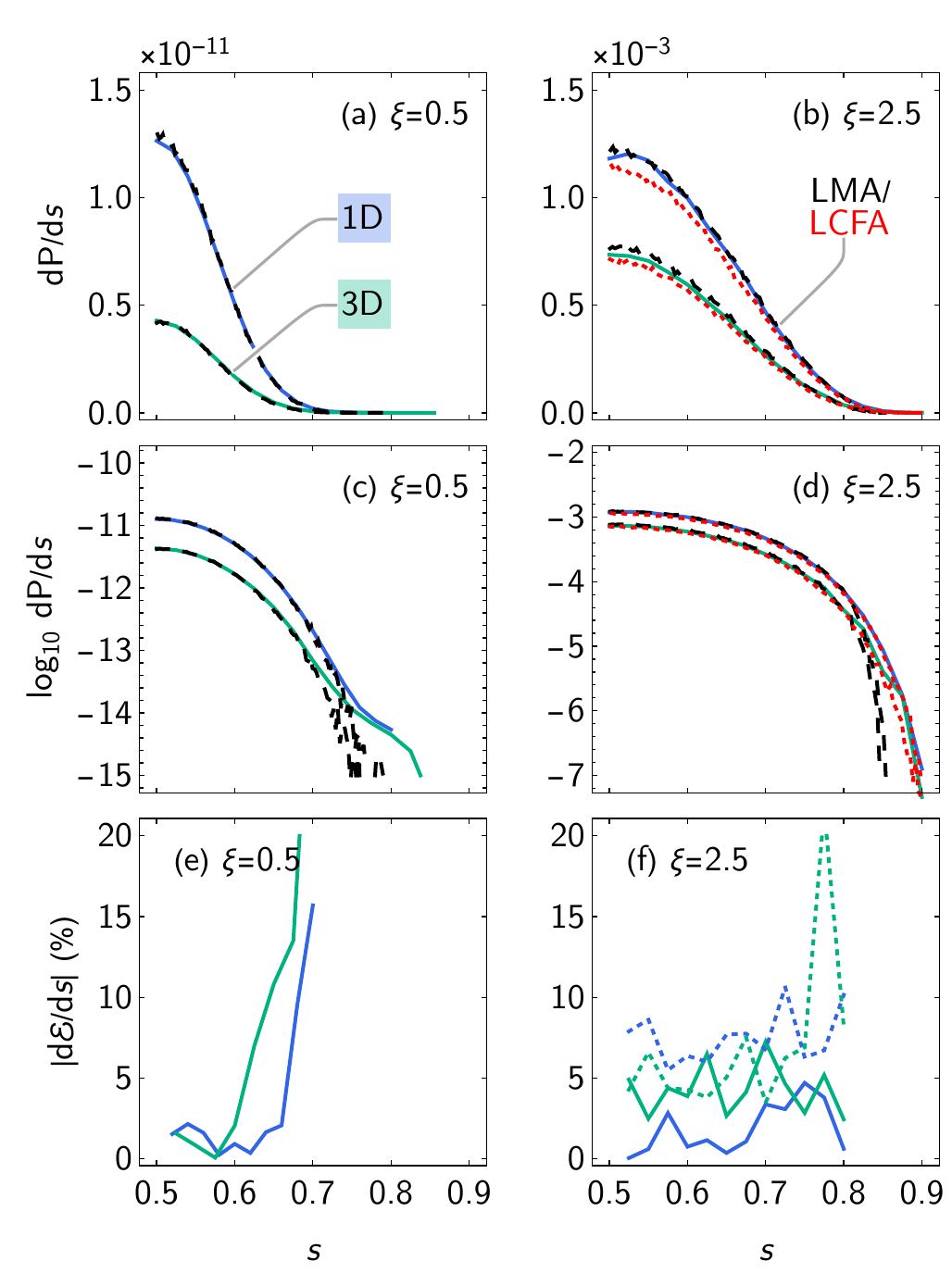}
    \caption{%
        Lightfront momentum spectra $\rmd\Prob/\rmd s$ for a photon with energy parameter $\eta = 0.2$ colliding with a plane-wave (blue) or focused (green) laser pulse of peak amplitude $\xi = 0.5$ (left column) and $2.5$ (right column) and duration $N = 16$.
        In the focused case, the waist $w_0 = 5\lambda$ and the photon beam has radius $2 \lambda$.
        (a-d) Results from QED (solid), LMA-based simulations (black, dashed) and LCFA-based simulations (red, short-dashed).
        (e,f) Percentage error made by the LMA (solid) and LCFA (dashed) simulations, in the plane-wave (blue) and focused (green) cases.}
    \label{fig:3D}
    \end{figure}

As an example of the agreement between simulation and QED for the case of a focused background, a scenario is calculated where photons with energy parameter $\eta=0.2$, arranged in a flat disc of radius $2\,\lambda$, collide head-on with a paraxial Gaussian beam with $N=16$ and a waist of $w_{0}=5\lambda$. In \figref{fig:3D}, agreement in the energy spectrum, $d\tsf{P}/ds$ is illustrated for two values of intensity, $\xi=0.5$ and $\xi=2.5$. 

In the multi-photon regime of $\xi=0.5$ (first column of \figref{fig:3D}), the centre of the spectrum ($s=1/2$) agrees with the QED result to give an error $d\tsf{P}/ds\approx 3\%  (\approx 1/2N)$. However, moving away from the centre of the spectrum, the disagreement between theory and simulation increases dramatically. One reason for this, which can be seen in the logarithmic plot (middle row), is that the QED result includes the linear Breit-Wheeler signal, which, being due to interference on the length scale of the pulse envelope, is missed by the LMA. The linear Breit-Wheeler signal has the effect of suppressing the rapid decay of the spectrum at larger values of $s$ \cite{king21dd}. This disagreement between simulation and theory, whilst large, is occurring in a region which does not contribute significantly to the total probability of the process.

In the all-order regime of $\xi=2.5$ (second column of \figref{fig:3D}), there is virtually no contribution from pulse-envelope interference, and the agreement between simulation and QED is good. For the 1D (plane wave) case, the agreement is $d\tsf{P}/ds\approx 3\%  (\approx 1/2N)$, and for the focused case, the error is slightly larger, of the order of $5\%$. (The comparison is made up to $s=0.8$ due to the finite number of harmonics included in the simulation rate, which leads to the rapid decay of the LMA away from the QED result, which can be seen in the logarithmic plot of the middle panel of the figure.) For comparison, the LCFA is also plotted, which performs worse in the 1D (plane-wave) comparison, but has a comparable error to the LMA for the 3D focused case at $\xi=2.5$.

\section{Conclusion}
The process of nonlinear Breit-Wheeler pair creation exhibits a perturbative, but generally nonlinear (multiphoton), dependency on the field strength when the intensity parameter $\xi\ll1$, but an all-order (non-perturbative) dependency on the field strength when $\xi\gtrsim 1$. To model this process for experiments using a field with intermediate intensity $\xi \sim O(1)$, a simulation framework is required that is accurate across these regimes. The standard method of including strong-field QED effects in numerical simulations, by employing the locally-constant field approximation (LCFA), is insufficient for this purpose, as the errors of the LCFA grow rapidly as $\xi$ is reduced from large values to values where $\xi \sim O(1)$ and smaller.

By directly comparing with numerical evaluation of exact QED expressions, the accuracy of simulations employing the locally monochromatic approximation (LMA) in calculating a range of observables and in different field configurations has been assessed. The accuracy can be quantified using an error function, $\mathcal{E}$, which is defined to be the relative difference between the QED and simulation results, and has the same functional dependence on parameters as the full or differential probability. The error function is useful for defining a `theory error' on simulation  results. 

Typically, in prediction of the total yield of pairs, the lightfront (energy) spectrum and the angular spectrum, $\mathcal{E}$ took values of the order of $1/\Phi$ (where $\Phi$ is the pulse duration), which is the expansion parameter of the LMA. Errors increase around harmonic channel openings, particularly around the $n=1$ and $n=2$ harmonics (which, for optical laser frequencies, require a probe photon energy of at least around $80\,\trm{GeV}$ and $40\,\trm{GeV}$ respectively, to access). Errors also increase when pulse-length interference effects begin to dominate, which can happen when it becomes favourable to create pairs via the linear Breit-Wheeler process using photons from a wide pulse envelope bandwidth. This effect is increased by: i) having a pulse shape with a wide bandwidth (in this paper the envelope was a squared cosine multiplied by a flat-top); ii) having lower probe photon energies (this means the yield of pairs is suppressed more strongly when $\xi$ is reduced in the multi-photon regime). In experiment, pulse envelope interference effects are likely to be minimal, because higher energy photons from the laser pulse will not be transmitted by various optical elements. (The pulse envelope chosen in the current paper has a particularly wide bandwidth.) A further source of error was found in the wings of pair energy spectra. When $\xi$ is increased above $\xi=1$, an increasing number of harmonics must be calculated in order that the decay of the spectrum in the wings is correctly predicted. Since only a finite number of harmonics can be tabulated in the numerical calculation, inevitably there are parts of the wings where the error can be large. However, this error occurs in regions of the spectrum that are orders of magnitude smaller than in the centre. Furthermore, as $\xi$ is increased, the LCFA becomes more accurate, and could potentially be used in a hybrid approach (perhaps along with an improved LCFA approximation \cite{Ilderton:2018nws,DiPiazza:2018bfu,King:2019igt}), if resolution of these high-energy tails were crucial to an experiment. A comparison was also made for pair creation in a focused laser pulse, where the baseline error was found to increase to around $5\%$ in regions where pulse-envelope and harmonic sampling effects were negligible. 

Therefore, although parameters can be found that test the accuracy of simulations calculating strong-field QED effects using a locally monochromatic approach, these parameters are for laser pulses or probe particle energies that are currently beyond any experimental realisation.

One can compare the current study of calculating nonlinear Breit-Wheeler with a similar study \cite{Blackburn:2021rqm} of nonlinear Compton. Whereas for Compton, harmonic structure in particle spectra can be accessed at low intensities and easily achievable electron energies, for Breit-Wheeler, clearly discernible harmonic structure requires very high probe photon energies, beyond anything currently planned in experiment. At intensity parameters $\xi<1$, the LCFA is often cited as becoming inaccurate for Compton, because of the importance of the lowest harmonics, which are missed by the LCFA. However, for the Breit-Wheeler process at $\xi<1$, the LCFA is much less accurate, and for experimentally relevant parameters, the threshold harmonic order is rather large. Thus for Breit-Wheeler, the inaccuracy of the LCFA is not due to missing harmonic structure, but rather the influence on the total probability due to interference effects on the length scale of the laser wavelength.

To conclude, in calculating how the yield of pairs depends on the intensity parameter $\xi$ being varied from the multiphoton ($\xi \ll 1$) to the all-order ($\xi>1$) regime, simulations based on the LMA sustain an accuracy level of around  $\lesssim 5\%$. This demonstrates the suitability of the approach for modelling high-energy experiments such as LUXE \cite{Abramowicz:2021zja} at DESY and E320 at SLAC.

\begin{acknowledgements}
BK acknowledges support from the Engineering and Physical Sciences Research Council (EPSRC), Grant No. \linebreak EP/S010319/1.
Simulations were performed on resources provided by the Swedish National Infrastructure for Computing (SNIC) at the High Performance Computing Centre North (HPC2N), partially funded by the Swedish Research Council through grant agreement no. 2018-05973.
\end{acknowledgements}

\bibliographystyle{apsrev4-1}
\bibliography{current}

\end{document}